\documentclass[twocolumn,prd,showpacs,floatfix]{revtex4}
\usepackage{epsfig}

\begin{document}

\title{New analysis of the SN~1987A neutrinos with a flexible
spectral shape}

\author{Alessandro Mirizzi$^{1,2}$ and Georg G.~Raffelt$^{1}$}

\affiliation{
$^1$Max-Planck-Institut f\"ur Physik (Werner-Heisenberg-Institut),
F\"ohringer Ring 6, 80805 M\"unchen, Germany\\
$^2$Dipartimento di Fisica and Sezione INFN di Bari,
                Via Amendola 173, 70126 Bari, Italy}

\date{\today}

\begin{abstract}
 
We analyze the neutrino events from the supernova (SN) 1987A detected
by the Kamiokande~II (KII) and Irvine-Michigan-Brookhaven (IMB)
experiments.  For the time-integrated flux we assume a quasi-thermal
spectrum of the form $(E/E_0)^\alpha\,e^{-(\alpha+1)E/E_0}$ where
$\alpha$ plays the role of a spectral index.  This simple
representation not only allows one to fit the total energy $E_{\rm
tot}$ emitted in $\bar\nu_e$ and the average energy $\langle
E_{\bar\nu_e}\rangle$, but also accommodates a wide range of shapes,
notably anti-pinched spectra that are broader than a thermal
distribution.  We find that the pile-up of low-energy events near
threshold in KII forces the best-fit value for $\alpha$ to the lowest
value of any assumed prior range.  This applies to the KII events
alone as well as to a common analysis of the two data sets.  The
preference of the data for an ``unphysical'' spectral shape implies
that one can extract meaningful values for $\langle
E_{\bar\nu_e}\rangle$ and $E_{\rm tot}$ only if one fixes a prior
value for $\alpha$. The tension between the KII and IMB data sets and
theoretical expectations for $\langle E_{\bar\nu_e}\rangle$ is not
resolved by an anti-pinched spectrum.
\end{abstract}


\maketitle

\section{Introduction}                        \label{sec:introduction}

The neutrino observations of supernova (SN) 1987A \cite{Hirata:1987hu,
Hirata:1988ad, Bionta:1987qt, Bratton:1988ww, Alekseev:1987ej,
Alekseev:1988gp} have long been taken as a confirmation of the salient
features of our physical understanding of the core-collapse SN
phenomenon. At the same time, a detailed
interpretation~\cite{Bahcall:1987ua, Krauss:1987re, Janka89,
Kernan:1994kt, Jegerlehner:1996kx, Kachelriess:2000fe, Loredo:1988mk,
Loredo:2001rx} of the data is difficult because of a number of
``anomalies''~\cite{Raffelt:1996wa}.  In particular, the energy
spectra implied by the Kamiokande~II (KII) and
Irvine-Michigan-Brookhaven (IMB) events are barely consistent with
each other. Moreover, the average $\bar\nu_e$ energy implied by KII is
much lower than expected from numerical simulations. The energies
implied by IMB alone would be compatible with theoretical models, but
the results of a common KII and IMB analysis are difficult to square
with expectations.

In the presence of neutrino oscillations, the $\bar\nu_e$ flux
observed in a detector is a superposition of the $\bar\nu_e$ and
$\bar\nu_{\mu,\tau}$ fluxes produced at the source.
Therefore, the KII and IMB detectors may have observed different
fluxes because the SN~1987A neutrinos have traversed different
sections of the Earth so that matter effects can modify the oberved
superposition of source spectra~\cite{Lunardini:2000sw,
Lunardini:2004bj}.  However, it is expected that $\langle
E_{\bar\nu_{\mu,\tau}}\rangle>\langle E_{\bar\nu_e}\rangle$ so that
oscillations aggravate the tension between observed and expected
$\bar\nu_e$ energies~\cite{Lunardini:2000sw, Lunardini:2004bj,
Jegerlehner:1996kx}.  In any event, the differences between $\langle
E_{\bar\nu_{\mu,\tau}}\rangle$ and $\langle E_{\bar\nu_e}\rangle$ are
probably much smaller than had been thought
previously~\cite{Raffelt:2001kv, Keil:2002in,Raffelt:2003en} so that
the Earth matter effect is no longer expected to cause gross
modifications of the observed $\bar\nu_e$ spectrum. Of course, the
relatively subtle modifications caused by Earth matter effects can be
crucial for identifying the neutrino mass ordering from the
high-statistics neutrino signal of a future galactic~SN
\cite{Dighe:1999bi, Takahashi:2001ep, Takahashi:2001dc,
Minakata:2001cd, Dighe:2003be, Dighe:2003jg, Dighe:2003vm,
Lunardini:2003eh, Fogli:2004ff, Tomas:2004gr}.

Previous studies of the SN~1987A neutrinos usually assumed a thermal
spectrum and then extracted $\langle E_{\bar\nu_e}\rangle$ and the
overall flux from the individual detector signals or from a combined
analysis. One exception is the analysis of Janka and
Hillebrandt~\cite{Janka89} who assumed an effective Fermi-Dirac
distribution of the form
\begin{equation}
\label{eq:fermi}
\varphi(E)\propto \frac{E^2}{e^{E/T-\eta}+1}\,,
\end{equation}
where $T$ is an effective temperature and $\eta$ a degeneracy
parameter. In their maximum-likelihood analysis they allowed only for
positive values of $\eta$. With this prior they found a best-fit value
of $\eta=0$ for both the KII and IMB data sets, suggesting that the
data prefer the broadest allowed distribution compatible with the
prior range for $\eta$.  However, even allowing for negative values
for $\eta$ would not have changed these results because
$\eta\to-\infty$ corresponds to a Maxwell-Boltzmann spectrum,
differing only marginally from the Fermi-Dirac case with $\eta=0$.

The main purpose of our new study is to investigate if better internal
agreement of the SN~1987A data as well as better agreement between the
data and theoretical expectations can be achieved if a more flexible
representation of the spectral shape is assumed.  Numerical studies of
neutrino transport suggest that the instantaneous spectra are
``pinched,'' i.e.\ that they are narrower than a thermal
spectrum~\cite{Keil:2002in,Janka89}.  A pinched spectrum can be
represented by a Fermi-Dirac distribution with positive
$\eta$. However, the SN~1987A data measure the time-integrated
spectrum, i.e.\ a superposition of instantaneous spectra with varying
average energies. Therefore, the integrated spectrum may well be
broader, not narrower, than a thermal spectrum, i.e.\ it may well be
anti-pinched.

If the time-intergrated spectrum is quasi-thermal in the sense that it
rises from zero for low energies, reaches a maximum, and has a long
tail to high energies, then the simplest conceivable representation
is~\cite{Keil:2002in}
\begin{equation}\label{eq:alpha-spectrum}
\varphi(E)=\frac{1}{E_0}
\frac{(\alpha+1)^{(\alpha+1)}}{\Gamma(\alpha+1)}
\left(\frac{E}{E_0}\right)^\alpha 
\exp\left[-(\alpha+1)\frac{E}{E_0}\right]\,,
\end{equation}
where $\int\varphi(E)\,dE=1$ and $E_0$ is an energy
scale with the property $\langle E\rangle = E_0$.
The numerical parameter $\alpha$ controls the width of the
distribution,
\begin{equation}
\frac{\langle E^2\rangle-\langle E\rangle^2}{\langle E\rangle^2}
=\frac{1}{1+\alpha}\,.
\end{equation}
We note that $\alpha=2$ corresponds to a Maxwell-Boltzmann spectrum,
$\alpha>2$ to a pinched spectrum with suppressed high- and low-energy
tails, and $\alpha<2$ to an anti-pinched spectrum. For
$\alpha\to\infty$ the spectrum becomes $\delta(E-E_0)$.
Since the oscillations effect
on the detected $\bar\nu_e$ spectrum is  reasonably small 
(see Sec.~\ref{sec:introduction}),
it makes sense to fit directly
the effective ${\bar \nu_e}$ spectrum, \emph{after} the oscillations, with the 
distribution
 of
Eq.~(\ref{eq:alpha-spectrum}).

In Sec.~\ref{sec:likelihood} we present the SN~1987A neutrino data and
perform a new maximum-likelihood analysis.  A summary and conclusions
are given in Sec.~\ref{sec:discussion}.

\section{Maximum-Likelihood Analysis}
\label{sec:likelihood}

\subsection{SN~1987A Data}

\begin{figure}[b]
\centering
\epsfig{figure=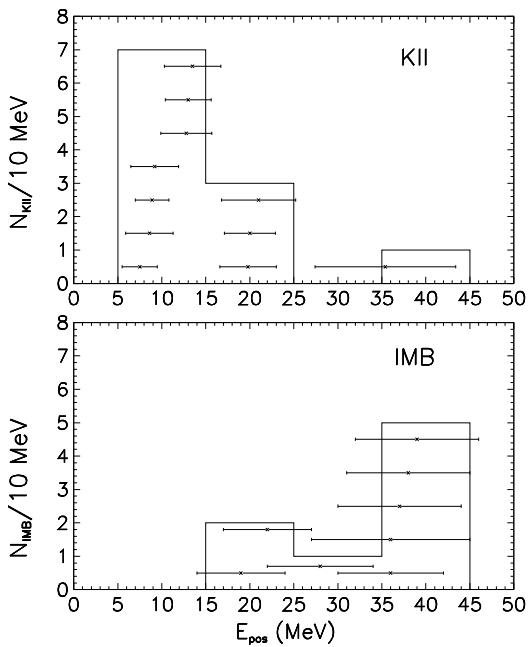,width =0.8\columnwidth,angle=0}
\caption{\label{fig:data} Positron spectra detected at KII (upper
panel) and IMB (lower panel) in connection with SN~1987A. We also show
the energies of the individual events together with their experimental
errors.}
\end{figure}

We limit our analysis to the SN~1987A data of the KII
\cite{Hirata:1987hu, Hirata:1988ad} and IMB \cite{Bionta:1987qt,
Bratton:1988ww} detectors that measure SN neutrinos almost exclusively
by the inverse beta reaction $\bar\nu_e+p\to n+e^+$. We show the
measured positron spectra in Fig.~\ref{fig:data} where we have left
out the KII event No.~6 that is attributed to background. In
Table~\ref{tab:data} we summarize the properties of the neutrino
signal in the two detectors. We do not include the Baksan Scintillator
Telescope (BST) data~\cite{Alekseev:1987ej, Alekseev:1988gp} because
it is much more uncertain which of the events have to be attributed to
background, i.e.\ in a maximum-likelihood analysis one would have to
model the background. This requires to use the time structure of the
neutrino burst~\cite{Loredo:1988mk, Loredo:2001rx}, whereas we limit
our study to the time-integrated flux.

\begin{table}[t]
\caption{Number of SN~1987A events and average positron energies at
the KII and IMB detectors.\label{tab:data}}
\begin{ruledtabular}
\begin{tabular}{lll}
Detector&$N_{\rm events}$&$\langle E_{e^+}\rangle$ [MeV]\\
\hline
KII & 11 & $15.4\pm1.1$ \\
IMB &  8 & $31.9\pm2.3$ \\
\end{tabular} 
\end{ruledtabular}
\end{table}

\subsection{Maxwell-Boltzmann Spectrum}

We perform a maximum-likelihood analysis of the SN~1987A signal along
the lines of the previous literature such as
Ref.~\cite{Jegerlehner:1996kx}. For the detection cross section we use
an updated analytic fit~\cite{Strumia:2003zx} and for the detection
efficiencies we use the analytic fit functions of
Ref.~\cite{Burrows:1988ba}.

\begin{figure}[ht]
\centering
\epsfig{figure=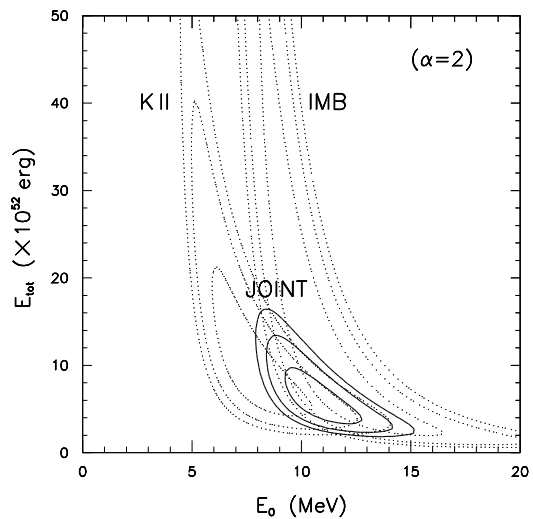,width =0.8\columnwidth,angle=0}
\caption{\label{fig2} Contours of constant likelihood that correspond
to 68.3\%, 95\% and 99\%\ C.L.\ in the plane $E_0$-$E_{\rm tot}$ for
an assumed Maxwell-Boltzmann ${\bar \nu_e}$ spectrum, corresponding to
$ \alpha=2$ in Eq.~(\ref{eq:alpha-spectrum}).  The dotted lines refer
to the KII and IMB signals, respectively, whereas the solid lines
represent a joint analyis.}
\end{figure}

In order to compare with the previous literature we first consider a
Maxwell-Boltzmann $\bar\nu_e$ spectrum, i.e.\ we use $\alpha=2$ in
Eq.~(\ref{eq:alpha-spectrum}). As fit parameters we use the average
$\bar\nu_e$ energy $E_0$ and the total energy $E_{\rm tot}$ emitted by
SN~1987A in the form of $\bar\nu_e$. Of course, these parameters refer
to the spectrum measured in the detectors after the partial flavor
swapping caused by neutrino oscillations. Our results shown in
Fig.~\ref{fig2} agree with the previous literature and illustrate once
more the tension between the average energies implied by the two
detectors.

\subsection{General Spectrum}

Next we use a general spectrum of the form
Eq.~(\ref{eq:alpha-spectrum}) with $E_0$, $E_{\rm tot}$, and $\alpha$
as our fit parameters with a prior $\alpha\geq0$. We show the best-fit
values for these parameters as well as the implied event numbers and
average positron energies in the detectors in
Table~\ref{tab:allfree}. This analysis is performed for each detector
separately as well as for the joint case.

\begin{table}[ht]
\caption{Best-fit values for $E_{\rm tot}$ in $10^{52}$~erg, $E_0$ in
MeV, and $\alpha$ based on the indicated data sets. The implied
characteristics of the expected detector signals are also
shown.\label{tab:allfree}}
\begin{ruledtabular}
\begin{tabular}{llllllll}
&\multicolumn{3}{l}{Best-fit param's}&
\multicolumn{2}{l}{KII}&
\multicolumn{2}{l}{IMB}\\
Data Set&$E_{\rm tot}$&$E_0$&
$\alpha$&$N_{\rm events}$&$\langle E_{e^+}\rangle$&
$N_{\rm events}$&$\langle E_{e^+}\rangle$\\
\hline
KII & 17.4   & 3.7  & 0.0  & 10.9 & 15.3 & 1.3 & 27.5  \\
IMB & 1.1 & 30.6 & 60.0  & 6.6 &30.3 & 7.9 & 31.4 \\
Joint & 10.1 & 5.4 & 0.0  & 14.1 & 19.2 & 4.9 & 32.0 \\
\end{tabular} 
\end{ruledtabular}
\end{table}

The IMB data alone prefer a functional form with a very large value
for $\alpha$, i.e.\ essentially a $\delta$ function.  This behavior is
intuitively obvious because 5 of the 8 IMB events have positron
energies in the very narrow range 36--39~MeV.  The KII data alone, on
the other hand, prefer $\alpha=0$, i.e.\ a huge total flux of
low-energy $\bar\nu_e$. The pile-up of events just above the KII
threshold of 7.5~MeV implies that the turn-over at low energies of a
quasi-thermal spectrum is not visible in the data, which are sensitive
only to the decreasing higher energy tail of the spectrum.  This
feature is reproduced in the likelihood by pushing $E_0$ toward low
values and broadening as much as possible the width.  Moreover, the
low value of $E_0$ is compensated by a huge amount of the total
emitted energy $E_{\rm tot}$.  We have checked that even when we allow
for negative $\alpha$ values the maximum of the likelihood always
coincides with the lowest allowed value. Of course, for $\alpha\leq-1$
the total neutrino flux diverges.

Contrary to our expectation, this behavior remains the same for a
joint analysis of the two data sets, even though the total flux is
smaller and the average $\bar\nu_e$ energy is larger. 

\begin{figure}[ht]
\centering
\epsfig{figure=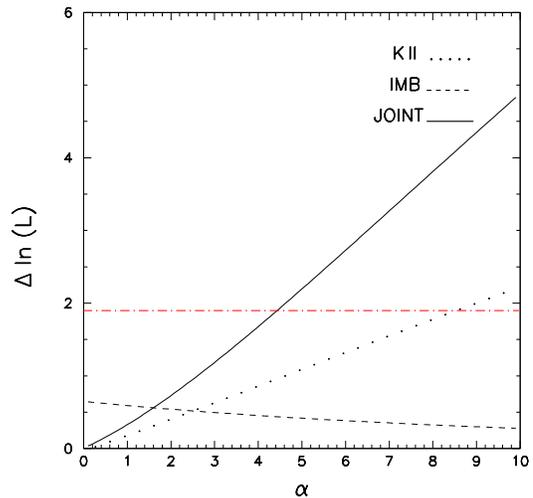,width =0.8\columnwidth,angle=0}
\caption{\label{fig3} Curves of $\Delta \ln {\mathcal L}$ relative to
the best-fit value as a function of $\alpha$ for KII (dotted line),
IMB (dashed line) and for the joint data set (continuous line). The
95\% C.L.\ for $\alpha$ is indicated with the dot-dashed horizontal
line. We have marginalized over $E_0$ and $E_{\rm tot}$.}
\end{figure}

In Fig.~\ref{fig3} we show the difference $\Delta\ln{\cal L}$ relative
to the best-fit value as a function of $\alpha$ for the two separate
data sets and the joint analysis. In all cases we have marginalized
over the parameters $E_0$ and $E_{\rm tot}$. The 95\% C.L.\ allowed
range for a single degree of freedom corresponds to
$\Delta\ln{\mathcal L}= 1.92$ shown as a horizontal dot-dashed line in
Fig.~\ref{fig3}. Evidently IMB alone has no strong preference for any
spectral shape. The 95\% upper limit on $\alpha$ is lower for the
combined data than for KII alone because the combined data also prefer
a larger $E_0$ and lower total flux. Overall, the data do not
distinguish in a useful way between different plausible spectral
shapes. In particular, the data do not prefer a quasi-thermal spectrum
but rather a monotonically falling one, in contrast with plausible
expectations.

\subsection{Fixed Prior Values for \boldmath{$\alpha$}}

\begin{table}[b]
\caption{Best-fit values for $E_{\rm tot}$ in $10^{52}$~erg and $E_0$
in MeV for the indicated fixed choices of $\alpha$.  The implied
characteristics of the expected detector signals are also
shown.\label{tab:fixed-alpha}}
\begin{ruledtabular}
\begin{tabular}{llllllll}
&&\multicolumn{2}{l}{Best-fit}&
\multicolumn{2}{l}{KII}&
\multicolumn{2}{l}{IMB}\\
&Data Set&$E_{\rm tot}$&$E_0$&
$N_{\rm events}$&$\langle E_{e^+}\rangle$&
$N_{\rm events}$&$\langle E_{e^+}\rangle$\\
\hline
$\alpha=0$
&KII   & 17.4   & 3.7    & 10.9 & 15.3 & 1.3 & 27.5  \\
&IMB   & 23.7 & 5.0   &28.7 &18.3 & 8.0 & 30.8  \\
&Joint & 10.1 & 5.4   & 14.1 & 19.2 & 4.9 & 32.0 \\[1ex]
$\alpha=2$
&KII& 8.9   & 7.9    & 11.0 & 15.3 & 1.1 & 26.7  \\
&IMB & 8.2 & 11.6   & 20.6 & 20.1 & 8.0 & 30.9  \\
&Joint & 5.9 & 11.2   & 14.0 & 19.5 & 4.9 & 30.3 \\[1ex]
$\alpha=4$
&KII & 6.6   & 10.2    & 11.0 & 15.4 & 1.0 & 25.3 \\ 
&IMB   & 4.7 & 15.9   & 16.3& 21.8& 8.0 & 31.0  \\
&Joint & 4.7 & 14.2  & 14.1 & 19.8 & 4.8 & 29.5 \\
\end{tabular}
\end{ruledtabular}
\end{table}

\begin{figure}
\centering
\epsfig{figure=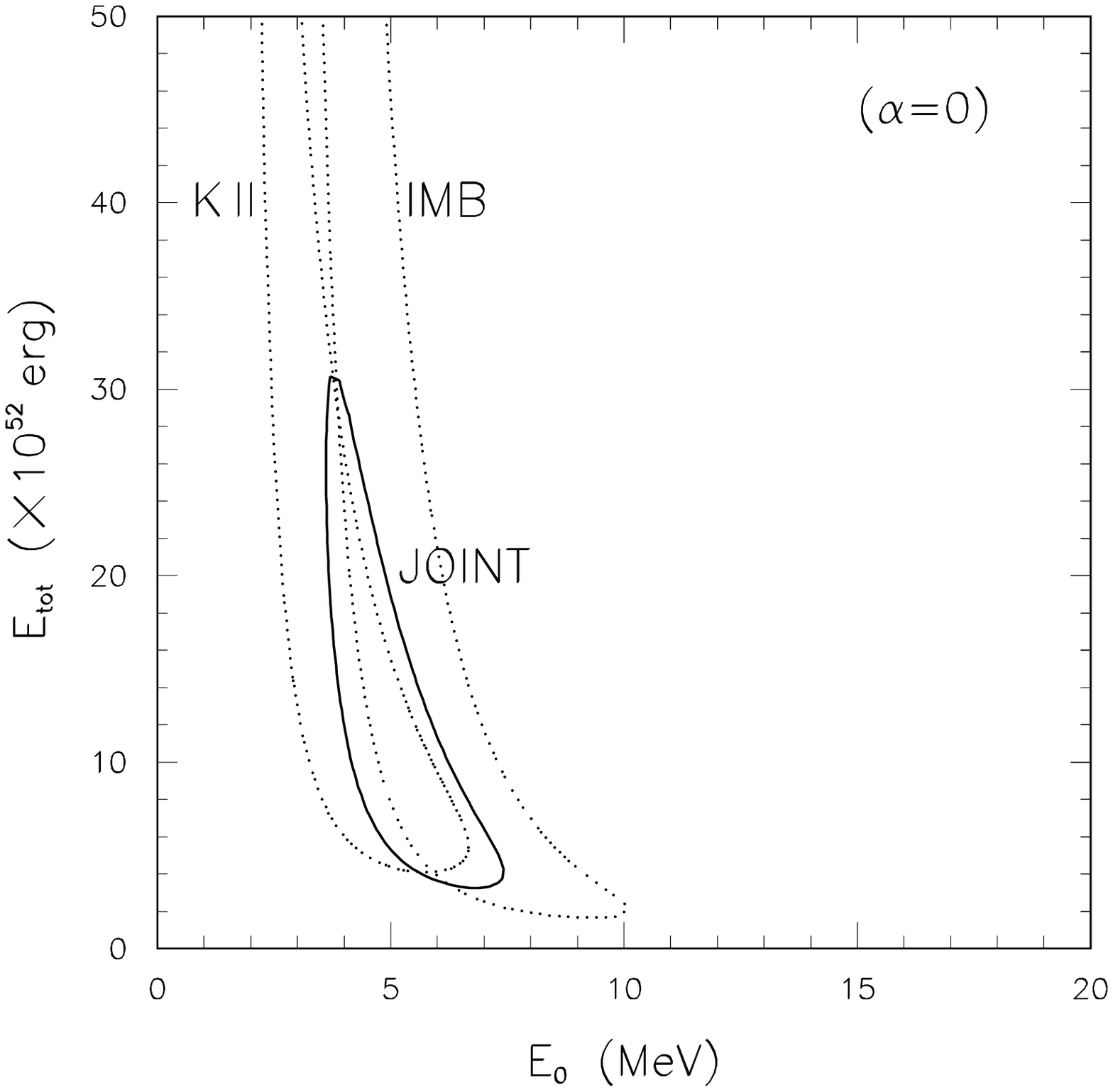,width =0.8\columnwidth,angle=0}
\vspace{0.5cm}
\epsfig{figure=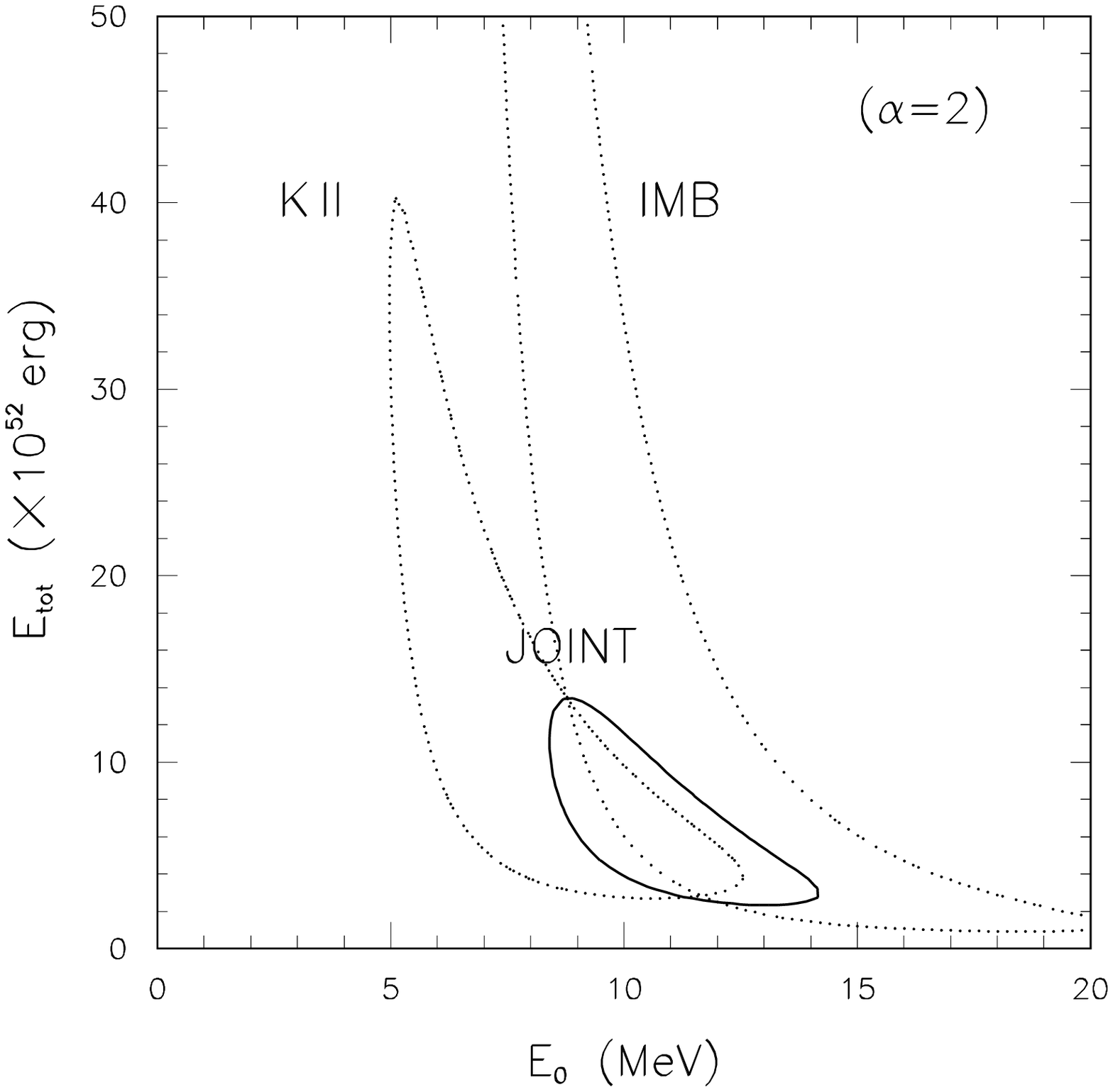,width =0.8\columnwidth,angle=0}
\vspace{0.5cm}
\epsfig{figure=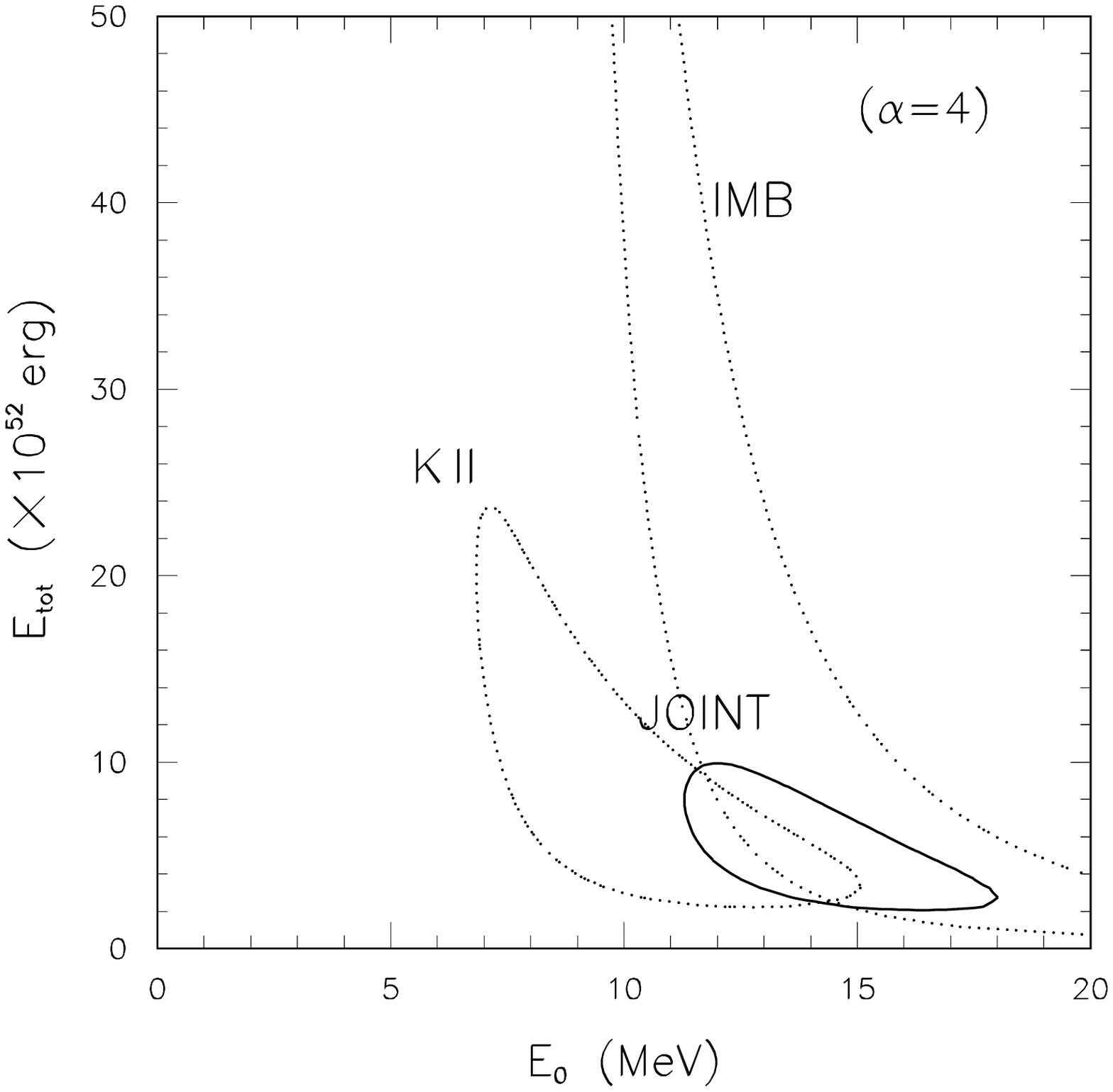,width =0.8\columnwidth,angle=0}
\caption{\label{fig4} Contours of 95 \% C.L.\ in the plane of $E_0$
and $E_{\rm tot}$ for $\alpha = 0$ (top panel), $ \alpha = 2 $
(middle) and $\alpha=4$ (bottom).  The dotted lines are for KII and
IMB separately whereas the solid line is for the joint data set.}
\end{figure}

We conclude that in order to extract information on the average
$\bar\nu_e$ energy and the total flux it is not useful to marginalize
over the parameter $\alpha$ because the result depends sensitively on
the assumed prior range for $\alpha$ and because unphysically low
values are prefered by the data. Therefore, we return to fixing the
spectral shape in advance and analyze the data for different choices
of $\alpha$.  In Fig.~\ref{fig4} we show the 95\% C.L.\ contours in
the $E_0$-$E_{\rm tot}$ plane for $\alpha=0$, 2 and 4 where the case
$\alpha=2$ is identical with the 95\% C.L.\ shown in
Fig.~\ref{fig2}. Larger values of $\alpha$ (pinched spectra) increase
the range of average energies and bring them closer to theoretical
expectations, but at the same time decrease the range of overlap
between IMB and KII.  In Table~\ref{tab:fixed-alpha} we show the
best-fit parameters $E_0$ and $E_{\rm tot}$ for the three $\alpha$
cases and different data sets as well as the implied signal
characteristics.

\section{Discussion and Summary}
\label{sec:discussion}

The expected neutrino energies from core-collapse SNe are of practical
importance for sensitivity forecasts of a future galactic SN neutrino
detection, in particular with regard to neutrino mixing
parameters~\cite{Dighe:1999bi, Takahashi:2001ep, Takahashi:2001dc,
Minakata:2001cd, Dighe:2003be, Dighe:2003jg, Dighe:2003vm,
Lunardini:2003eh, Fogli:2004ff, Tomas:2004gr}.  The expected energies
are also important in the context of current limits and future
detection possibilities of the diffuse SN neutrino background from all
SNe in the universe \cite{Malek:2002ns, Ando:2004hc, Strigari:2005hu,
Fogli:2004gy,Beacom:2005it}. The internal tension within the SN~1987A neutrino data
as well as the tension with theoretical expectations has persuaded
most workers in this field to ignore the data and rely on the output
of numerical simulations even though current SN theory may still be
missing an important piece of input physics to obtain robust
explosions~\cite{Buras:2003sn}.

One explanation for the tension between the KII and IMB data is that
the detectors actually observed different spectra due to the Earth
matter effect~\cite{Lunardini:2000sw, Lunardini:2004bj}. However, our
current understanding of flavor-dependent neutrino spectra formation
suggests that the flavor-dependent average energies in the
anti-neutrino sector are not very different~\cite{Raffelt:2001kv,
Keil:2002in}.  Moreover, if the observed $\bar\nu_e$ had been born as
higher-energy $\bar\nu_{\mu,\tau}$ at the source, the tension between
theoretically expected and actually observed $\bar\nu_e$ energies
would be worse.

Loredo and Lamb~\cite{Loredo:2001rx} have tested a large variety of
neutrino emission models and in particular have included what amounts
to a bi-modal spectral shape that contains a lower-energy component
attributed to the SN accretion phase and a higher-energy one
attributed to the neutron-star cooling phase.  They stress that such a
spectral form is strongly favored by the data relative to a
single-mode thermal spectrum.  However, numerical simulations do not
predict a bimodal spectrum because the average energies continuously
increase from the accretion to the neutron-star cooling phase and at
late times decrease again.  Therefore, the spectral shape of the
time-integrated flux does not seem to exhibit a bi-modal form but
rather is expected to be a broadened quasi-thermal spectrum.

Motivated by this observation we have analysed the SN~1987A data,
assuming a quasi-thermal spectrum of the general form
Eq.~(\ref{eq:alpha-spectrum}) that is flexible enough to accommodate a
continuum of spectral shapes from narrow (pinched) to broadened
(anti-pinched) spectra relative to a Maxwell-Boltzmann distribution.

Perhaps unsurprisingly in view of the tension between the KII and IMB
data, we find that the broadest possible distribution allowed by the
chosen prior range of $\alpha$ is prefered. In this way the tension
between the data sets is somewhat reduced, but at the same time the
average $\bar\nu_e$ energies are pulled to lower values, thus
exacerbating the tension with the output of representative numerical
simulations.  (The average neutrino energies found in many different
numerical simulations have been collected in Ref.~\cite{Keil:2002in}.)

Assuming the time-intergrated neutrino flux from a SN indeed exhibits
a quasi-thermal spectrum roughly of the form
Eq.~(\ref{eq:alpha-spectrum}), the implied average $\bar\nu_e$
energies and total emitted energy depend sensitively on the chosen
prior range for $\alpha$. The data themselves prefer the smallest
possible $\alpha$-values, i.e.\ not a quasi-thermal spectrum but
rather a monotonically falling one. The assumption of a realistic
quasi-thermal spectrum is not simultaneously consistent with typical
theoretical expectations for $\langle E_{\bar\nu_e}\rangle$ as well as
the separate data sets from IMB and KII. Therefore, it remains
unresolved if the SN~1987A data or theoretical simulations give us
better benchmarks for the average $\bar\nu_e$ energies to be used, for
example, in the context of searches for the cosmic diffuse SN neutrino
background. 

It appears that the question of the true neutrino spectrum from a
typical SN can be empirically resolved only by the high-statistics
signal from a future galactic SN or by the patient accumulation of
data on a neutrino-by-neutrino basis from SNe in nearby
galaxies~\cite{Ando:2005ka}.


\begin{acknowledgments}
We acknowledges partial support by the Deutsche Forschungsgemeinschaft
under Grant No.~SFB-375 and by the European Union under the Ilias
project, contract No.~RII3-CT-2004-506222.  The work of A.M.\ is
supported in part by the Italian ``Istituto Nazionale di Fisica
Nucleare'' (INFN) and by the ``Ministero dell'Istruzione, Universit\`a
e Ricerca'' (MIUR) through the ``Astroparticle Physics'' research
project.
\end{acknowledgments}


\end{document}